\documentclass[aps,prl,twocolumn,noshowpacs,superscriptaddress,groupedaddress]{revtex4}  
\usepackage{pdfpages}
\usepackage{xcolor}

\usepackage{graphicx}  
\usepackage{dcolumn}   
\usepackage{bm}        
\usepackage{amssymb}   
\usepackage{amsmath}
\usepackage{color}
\usepackage[colorlinks=true, pdfstartview=FitV, linkcolor=blue, citecolor=blue, urlcolor=blue]{hyperref} 
\usepackage{epstopdf}
\usepackage{todonotes} 
\usepackage{lipsum}  
\usepackage{appendix} 
\usepackage{siunitx}

\usepackage{tabularx} 
\usepackage{xspace}
\usepackage{braket}
\usepackage{color}
\usepackage{setspace}
\usepackage{mdframed}
\usepackage{framed}
\usepackage{multirow}
\usepackage{setspace}
\usepackage[export]{adjustbox}
\usepackage{varwidth}
\usepackage{float}

\newcommand{\ee}[1]{\times 10^{#1}}

\newcommand{\SF}{S_\mathrm{th}}

\newcommand{\affilETH}{Laboratory for Solid State Physics, ETH Z\"{u}rich, 8093 Z\"urich, Switzerland.}
\newcommand{\affilNBI}{Niels Bohr Institute, University of Copenhagen, 2100 Copenhagen, Denmark.}
\newcommand{\affilNBIHYQ}{Center for Hybrid Quantum Networks, Niels Bohr Institute, University of Copenhagen, 2100 Copenhagen, Denmark.}
\newcommand{\affilUnibas}{Department of Physics, University  of  Basel, 4056  Basel, Switzerland.}
\newcommand{\affilGroningen}{Department of Biomedical Engineering, Groningen University, 9713 AW Groningen, The Netherlands.}
\newcommand{\affilChicago}{Present address: Pritzker School of Molecular Engineering, University of Chicago, Chicago, IL 60637, USA.}

\begin{document}

\author{David H\"alg}
\thanks{These authors contributed equally to this work.}
\affiliation{\affilETH}
\author{Thomas Gisler}
\thanks{These authors contributed equally to this work.}
\affiliation{\affilETH}
\author{Yeghishe Tsaturyan}
\affiliation{\affilNBI}
\affiliation{\affilChicago}
\author{Letizia Catalini}
\affiliation{\affilNBI}
\affiliation{\affilNBIHYQ}
\author{Urs Grob}
\affiliation{\affilETH}
\author{Marc-Dominik Krass}
\affiliation{\affilETH}
\author{Martin H\'eritier}
\affiliation{\affilETH}
\author{Hinrich Mattiat}
\affiliation{\affilUnibas}
\author{Ann-Katrin Thamm}
\affiliation{\affilETH}
\author{Romana Schirhagl}
\affiliation{\affilGroningen}
\author{Eric C. Langman}
\affiliation{\affilNBI}
\affiliation{\affilNBIHYQ}
\author{Albert Schliesser}
\affiliation{\affilNBI}
\affiliation{\affilNBIHYQ}
\author{Christian L. Degen}
\affiliation{\affilETH}
\author{Alexander Eichler}
\affiliation{\affilETH}

\title{Membrane-based scanning force microscopy}


\preprint{} 

\date{\today} 

\begin{abstract}
We report the development of a scanning force microscope based on an ultra-sensitive silicon nitride membrane transducer. Our development is made possible by inverting the standard microscope geometry -- in our instrument, the substrate is vibrating and the scanning tip is at rest. We present first topography images of samples placed on the membrane surface. Our measurements demonstrate that the membrane retains an excellent force sensitivity when loaded with samples and in the presence of a scanning tip. We discuss the prospects and limitations of our instrument as a quantum-limited force sensor and imaging tool.

\end{abstract}


\maketitle

\section{Introduction} 

The coupling between optical and mechanical degrees of freedom plays a fundamental role in physics and forms the basis for many important applications in metrology and signal transduction~\cite{Aspelmeyer_2014}.
The field of cavity optomechanics, in particular, has enabled the detection of vibrations with remarkably high sensitivity~\cite{Teufel_2009, Anetsberger_2009} extending to below the standard quantum limit~\cite{mason_continuous_2019}, demonstrated squeezed phononic states~\cite{Wollman_2015, Pirkkalainen_2015, Lecocq_2015} and quantum ground state cooling of `massive' mechanical resonators~\cite{Wilson_2004, Wilson_2007, Teufel_2011, Chan_2011, Peterson_2016}, and enabled optoelectromechanical signal transduction with quantum-limited efficiency~\cite{Bagci_2014, Andrews_2014}. Such advances raise the question whether the extreme sensitivity of optomechanical systems could be harnessed for a new generation of quantum-limited force sensors and scanning force microscopes~\cite{Milburn_1994}. 

        \begin{figure}
        \includegraphics[width=1.02\columnwidth]{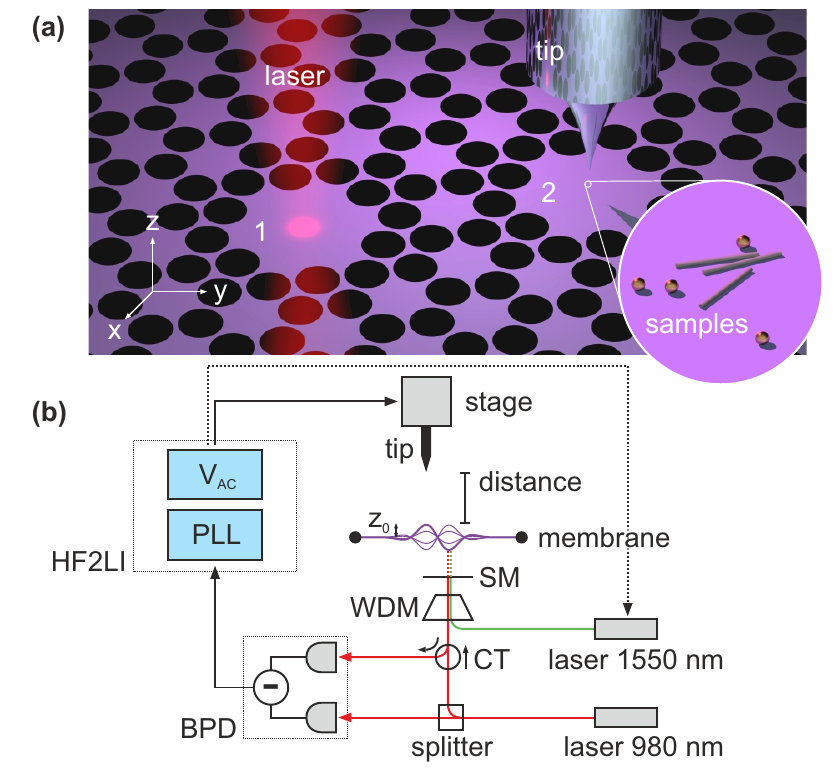} 
        \caption{\textbf{Configuration of the inverted scanning force microscope.} (a) We use a $41$-nm-thick silicon nitride membrane as nanomechanical force sensor. A hole pattern acts as a phonon shield, while two `defect' areas (labeled 1 and 2) undergo out-of-plane vibrations. The areas are coupled and form hybridized normal modes. We load gold nanoparticles and tobacco mosaic viruses onto one of the defect areas and measure the vibrations on the second one with a laser interferometer. A metallic scanning tip interacts with the samples and modifies the vibrations. The separation of the defect centers is 530~$\mu$m.  The scanning force microscope is operated in high vacuum to avoid viscous damping.  Figure adapted from Ref. \citenum{Kosata_2020}.
        (b) A first 1550~nm laser is used to detect the membrane motion and a second 980~nm laser is used to drive the membrane via radiation-pressure forces.  Membrane vibrations are detected by feeding the photodetector signal into a lock-in amplifier ({\sc HF2LI, Zurich Instruments}). The ac output voltage of the lock-in amplifier is used to modulate the potential of the scanning tip or the power of the driving laser. CT: circulator. WDM: wavelength-division multiplexer. SM: semitransparent mirror. BPD: balanced photodetector. PLL: phase-locked loop.
        }
        \label{fig:fig1}
        \end{figure}

Sensitive force microscopy has important applications in the detection of weak electric and magnetic phenomena. Examples include the imaging of isolated electronic charges~\cite{Klein_2001, Stomp2005} and spins~\cite{Rugar_2004}, nanoscale nuclear spin density distributions~\cite{Poggio_2007,Degen_2009}, as well as magnetic fields generated by nanoscale currents~\cite{Ando2000}. 
A long-standing goal is the detection of a single nuclear spin, which would constitute a milestone towards atomic-resolution magnetic resonance imaging~\cite{Sidles_1991, Degen_2009, Mamin_2013, Staudacher_2013, Loretz_2014}. This vision hinges crucially on the development of mechanical transducers that combine exceptionally low thermomechanical force noise with very low detector noise. In recent years, impressive progress in both respects has been achieved with silicon nitride (Si$_3$N$_4$) strings and membranes, whose high strain, extreme aspect ratio, and phononic crystal clamp engineering have resulted in very high quality factors~\cite{Verbridge_2008, Zwickl_2008, Unterreithmeier_2010, Schmid_2011,  Yu_2012, Chakram_2014, Reinhardt_2016, Norte_2016, Tsaturyan_2017, Rossi_2018, Ghadimi_2018}.
Membranes and string resonators, however, require clamping on opposite ends and can therefore not be used in the cantilever or pendulum geometry adopted in ultrasensitive scanning force microscopy experiments~\cite{Scozzaro_2016, Fischer_2019}. This obstacle has so far impeded the implementation of membranes in nanoscale scanning force microscopy.

In this work, we demonstrate a scanning force microscope based on a silicon nitride membrane sensor. We accommodate the seemingly incompatible geometries by inverting the standard microscope paradigm; in our setup, the sample plate is also the force sensor, which we approach with a static (non-vibrating) scanning tip.
We avoid snap-in-to contact by employing high-stiffness membranes,
and achieve high force sensitivity through the use of state-of-the-art phononic bandgap membranes~\cite{Tsaturyan_2017} with very high quality factors ($Q=10^7-10^8$).
Indeed, we find that the membrane transducers preserve their quality factors and force sensitivities for tip-surface distances as small as a few nm.
At the same time, the extended membrane surface allows placement of a wide range of samples.  We demonstrate scanning probe operation by imaging the nanometer-resolution topography of gold nanospheres and tobacco mosaic virus (TMV) particles deposited on the membrane surface.
When combined with a precise optical cavity readout and novel spin-mechanics coupling protocols~\cite{Kosata_2020} at cryogenic temperatures, our approach is expected to enable quantum-limited scanning force microscopy with the capability to detect individual nuclear spins.

\section{Device and Setup}

      \begin{figure*}
        \includegraphics[width=\textwidth]{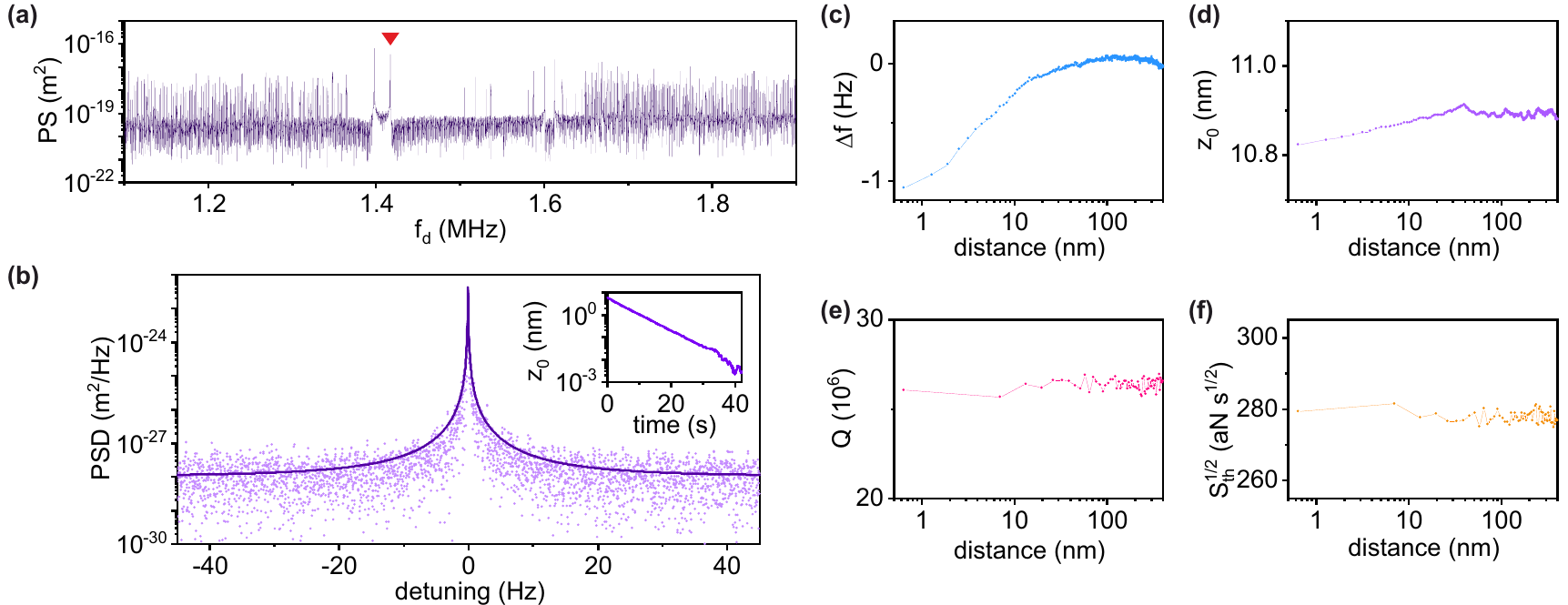} 
        \caption{\textbf{Membrane properties and tip approach.} (a) Displacement power spectrum (PS) obtained by slowly sweeping $f_d$ and measuring the membrane response with a lock-in amplifier. The red arrow marks the high-$Q$ mode used in this work. The force was applied through the modulated potential of the scanning tip far away from the surface. (b) Thermal displacement power spectral density (PSD) of the high-$Q$ mode. The solid line is a fit to the model of a harmonic resonator using $Q = 2.6\times 10^7$, $f_0 = 1.417~$MHz and yielding an effective mass of $m = 14~$ng.
        The spring constant is $k = m(2\pi f_0)^2 = 1100~$N/m. The inset shows a ringdown measurement to confirm the value of $Q$. (c)-(e) Shift of the mechanical resonance frequency, vibration amplitude, and quality factor measured as a function of tip-surface distance, respectively. The membrane is driven by a radiation pressure force with a constant amplitude.
        We define the touch position ($d=0$) from threshold values where $\Delta f$ and $z_0$ exceed a set value. (\textit{e.g.} $\Delta f=0.3$~Hz and $\Delta z_0=30$~\% for the scans shown in this paper.)
        (f) Intrinsic single-sided thermomechanical force noise calculated from (c)-(e) and the fit results from (b).
        All measurements in the main manuscript are from a single membrane device. Data from additional devices are shown in the Supplemental Information (SI).
        }
        \label{fig:fig2}
        \end{figure*}
        
Our inverted scanning force microscope consists of a static tip scanning above a vibrating silicon nitride membrane (Fig.~\ref{fig:fig1}).  The membrane oscillates in the out-of-plane direction, unlike other ultrasensitive force microscopes that use a pendulum geometry~\cite{Heritier_2018}.
The membrane simultaneously acts as the sample platform and the optomechanical force transducer.  To improve the quality factor and hence the force sensitivity, we use membranes that have been structured with a hexagonal lattice of perforations to create a bandgap for out-of-plane oscillations.  At the center of the membrane sits a larger ($\sim 100\,\mathrm{\mu m}$) unperforated area.  This `defect' is capable of supporting localized mechanical modes with eigenfrequencies inside the bandgap and mechanical quality factors up to one billion~\cite{Tsaturyan_2017, Rossi_2018}.  Similarly high quality factors have been reported for other types of perforations~\cite{Regal2019} and for geometries with narrow tethers~\cite{Sankey2017}.

In our experiments, we work with membranes that have a pair of defects.  The defects are separated by a distance of a few hundred $\mu$m such that they are mechanically coupled and form hybridized normal modes  -- a symmetric and an anti-symmetric mode -- with frequencies that are a few tens of kHz apart (see Fig. \ref{fig:fig1}a)~\cite{Catalini_2020}.
In future experiments, we expect to use the double-mode structures for parametric up-conversion of magnetic forces from nuclear spins~\cite{Kosata_2020}. 
In the present work we simply take advantage of the fact that the two defects allow us to spatially separate the sites for sample placement and for optical readout.

We detect the membrane vibrations with a fiber-based laser interferometer positioned below the membrane (Fig.~\ref{fig:fig1}(b))~\cite{Rugar1989}. The two arms of the interferometer are formed by a semitransparent mirror at the fiber end and by the membrane itself.  Due to the low membrane reflectivity ($\sim 6\%$) we employ a balanced photo-detector that offsets the large reflection from the reference arm.
The interferometer has a double-sided power spectral density (PSD) noise floor of $\sim 10^{-28}~\rm{m}^2/\sqrt{\rm{Hz}}$. We drive the membrane either via the radiation pressure force of a second driving laser or through an electrostatic modulation of the scanning tip (see Fig. \ref{fig:fig1}(b)).
We detect the signal by a lock-in amplifier, and utilize the amplifier's on-board phase-locked loop (PLL) to maintain resonant driving of membrane modes in the presence of thermal frequency drifts.

To realize the scanning probe functionality, we integrate a scanning tip above the membrane.  The tip is mounted on an XYZ piezo translation stage and approached vertically. For the measurements presented here, we use metal-coated silicon tips from commercial atomic force microscope (AFM) cantilevers.  The tips are glued to a micro-machined metallic needle acting as a rigid support. The metal finish of the scanning tip is convenient for applying a tip potential and for implementing a dynamic electrostatic driving.  We have also tested scanning tunneling microscopy tips made from tungsten by wet etching, but found them to degrade rapidly due to the high stiffness of the membrane.

\section{Tip Approach}

Before demonstrating scanning experiments, we present a few defining features of the double-defect membrane resonators.
To characterize a membrane, we first sweep the frequency $f_d$ of a small force to drive vibrations in a spectral range of interest.
The frequency spectrum shows a dense quasi-continuum of modes that is interrupted by a gap from $\sim 1.36~$MHz to $1.65~$MHz (Fig.~\ref{fig:fig2}(a)). Inside the gap, we find only a few out-of-plane modes that typically feature quality factors $Q>10^7$~\cite{Tsaturyan_2017}.  We identify the double-defect modes by the pair of resonances around $1.4~$MHz whose frequency splitting agrees with that expected from theory~\cite{Catalini_2020}. In the following, we focus on the anti-symmetric mode with a resonance frequency of $f_0 = 1.417~$MHz. We determine the precise quality factor of the modes via the thermomechanical displacement noise spectrum and through ring-down measurements (Fig. \ref{fig:fig2}(b)).


To engage the tip, we approach the surface of the membrane until we observe a shift of the resonance frequency $\Delta f$ as well as a reduction of the amplitude $z_0$ (see Fig.~\ref{fig:fig2}(c-d)).
We find that the frequency and amplitude only change at very close approach, $d<10~$nm.  This behavior is expected because the membrane is under strong tension and has a high stiffness $k\sim10^3~$N/m.
%
We further measure the mechanical quality factor $Q$ as a function of distance, and observe no noticeable decrease even at very close approach (Fig.~\ref{fig:fig2}(e)).  This observation is important, because it indicates little non-contact friction~\cite{Kuehn_2006, Kisiel_2011} and no loss in force sensitivity in close proximity to the membrane. 
Shown in Fig.~\ref{fig:fig2}(f), the single-sided thermomechanical force noise, defined by $S_\mathrm{th}^{1/2}=4k_\mathrm{B}T \gamma_0$, is approximately $280~$aN/$\sqrt{\mathrm{Hz}}$ at a distance of $20~$nm and changes by less than 1\% when lowering the distance to $1~$nm. Here, $T \approx 300~$K is the ambient temperature, $k_B$ is Boltzmann's constant, and $\gamma_0 = 2\pi f_0 m/Q$ is the damping coefficient when the scanning tip is far from the surface, with $m = 14~$ng the effective mass. The damping added by non-contact friction at $d=1~$nm thus only amounts to $\gamma_{\mathrm{nc}} \lesssim 0.01 \gamma_0 \approx 5.5\times 10^{-14}~$kg/s. This value compares favorably to $\gamma_{\mathrm{nc}} \approx 1.5\times 10^{-13}~$kg/s and $ 10^{-12}~$kg/s found at room temperature for a pendulum-style cantilever over a gold surface at distances of $10$ and $2~$nm, respectively~\cite{Stipe_2001}. The low non-contact friction of our membrane sensor is probably be due to a combination of (i) reduced electrical fluctuations at MHz frequencies relative to the kHz range~\cite{Braakman_2019, Labaziewicz2008}, and (ii) smaller interaction between tip and surface for out-of-plane oscillations compared to the shear motion in the pendulum geometry. We expect the non-contact friction to be further reduced at cryogenic temperatures~\cite{Stipe_2001,Nichol_2012}.


    \begin{figure*} 
    \includegraphics[width=170mm]{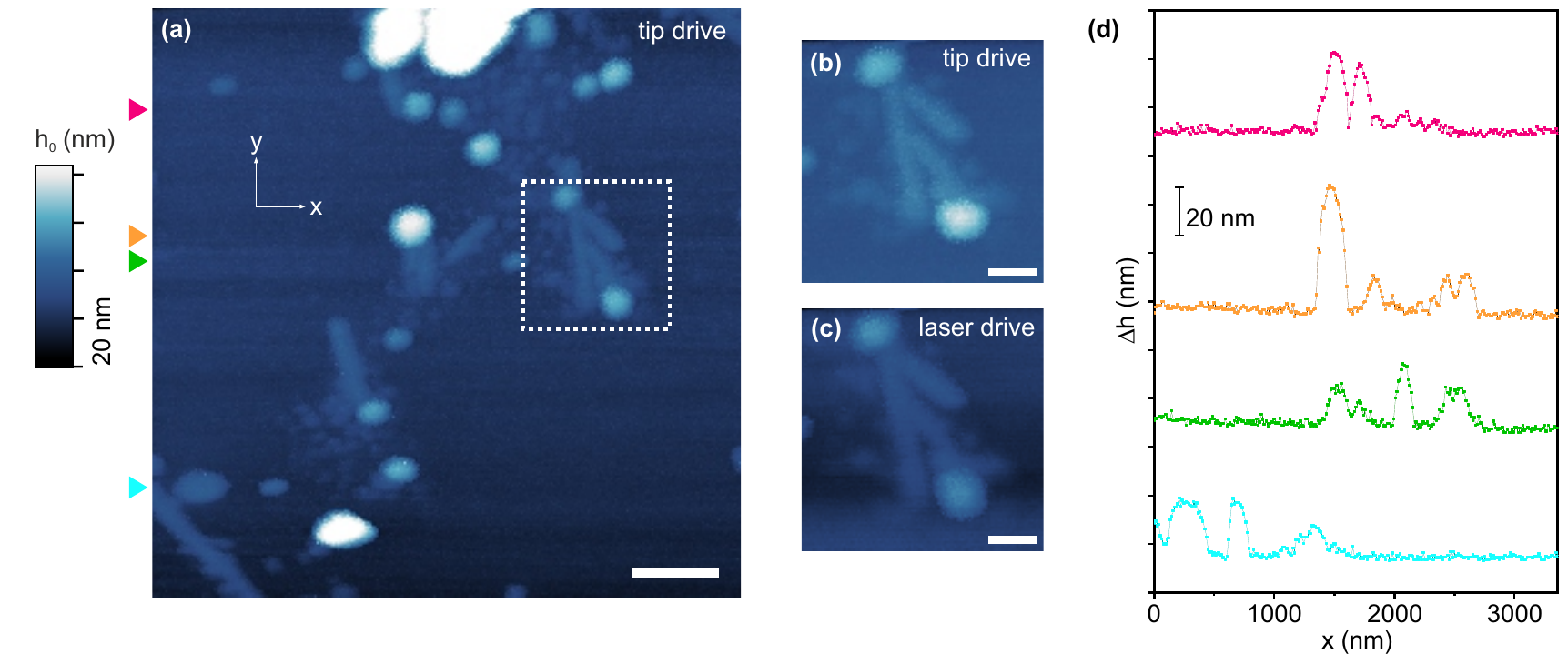} 
    \caption{\textbf{Topographic imaging}. The relative surface position $h_0$ is indicated by the same color bar for all images, with ticks separated by $20~$nm. (a) Touch map of membrane surface with gold nanoparticles (nominal average diameter of $50~$nm) and TMV samples (diameter of $18~$nm, used for $z$-scanner calibration~\cite{Trinh2011}). Lateral scale bar is $500~$nm and step size is $13~$nm. Marks on the left side indicate the positions of the linescans in (d). The membrane is driven electrostatically by modulating the tip potential. (b) High-resolution scan over the area indicated by a square in (a). Lateral scale bar is $200~$nm and step size is $10~$nm. (c) Same scan performed using radiation pressure driving of the membrane. Lateral scale bar is $200~$nm and step size is $10~$nm. (d) Linescans extracted from the data in (a). Scans are vertically offset for clarity.}
    \label{fig:fig3}
    \end{figure*}

\section{Topographic Imaging}


We demonstrate scanning force microscopy by imaging the topography of samples dispersed on the membrane surface. Because of the high stiffness of the membrane, we build up images by measuring the `touch point' at each pixel, rather than by regulating the frequency or amplitude via distance feedback.
We define the touch position from threshold values of $\Delta f$ and $z_0$, that is, we terminate the approach (and define a `touch') when the change in either frequency or amplitude exceeds a set value.  As shown below, this procedure is reproducible with sub-nanometer accuracy.
In Fig.~\ref{fig:fig3}(a), we show a topographic image of gold nanospheres and tobacco mosaic virus (TMV) particles that are deposited from liquid suspension (see SI for details). Round and elongated objects are clearly visible and correspond in height and length roughly to the expected dimensions of the gold and TMV particles, respectively. Zoom scans show the same smaller area, once with the membrane driven by an electrostatic force applied through the tip (Fig.~\ref{fig:fig3}(b)), and once by radiation pressure driving (Fig.~\ref{fig:fig3}(c)).  The two images are nearly identical and demonstrate that both actuation methods are equally well suited for scanning imaging.

In order to evaluate the spatial resolution, we extract linescans in $x$-direction at different $y$-positions (Fig.~\ref{fig:fig3}(d)). The step size is $13~$nm in these scans, and we find dips between objects that narrow down to a single point. 
From the flat part of the lowest linescan between $2000$ and $3360~$nm, we obtain a standard deviation in z-direction of $0.3~$nm.  This value is an upper bound on the resolution in height, as we cannot exclude that some fluctuations stem from real features.  The lateral resolution is lower due to convolution with the rounded AFM tip.  Given that the cylindrical TMV features in our image have lateral widths of $\sim 100~$nm, we estimate that the scanning tip has a nominal radius of about $40~$nm.  If desired, this resolution could be improved by integrating high-resolution AFM probes.


Our touch map imaging method currently is very slow.  Each pixel requires roughly $8~$s of acquisition time, and the measurement of a large and detailed image, such as the one shown in Fig.~\ref{fig:fig3}(a) with $256\times 256$ pixels, requires several days.  This limitation is mainly due to the high stiffness of the membrane, which prevents the use of standard $z$-feedback regulation and is ill suited for the imaging of large ($>10~$nm) topographic features.  We expect that this issue will be alleviated with a next generation of membranes with a lower stiffness in the $1-10~$N/m range (SI).


\section{Discussion and Outlook}

An important prospect of optomechanical membranes is their potential for enabling quantum-limited, sub-attonewton force microscopy.
We conclude our paper by discussing the challenges and key steps towards reaching this goal.
The membrane used in this work has an intrinsic single-sided  room-temperature thermomechanical force noise of $S_{\rm{th}}^{1/2} = 280~\rm{aN\,Hz}^{-1/2}$ (Fig.~\ref{fig:fig2}(f)). This figure is not yet competitive with state-of-the-art cantilevers~\cite{Heritier_2018}, carbon nanotubes~\cite{Moser_2013, deBonis_2018} or nanowires~\cite{Rossi_2017, Nichol_2012, Delepinay_2017, Sahafi_2019} that featured measured force noise on the order of $10$ to $100~\rm{aN\,Hz}^{-1/2}$ at room temperature, and that reached $<1~\rm{aN\,Hz}^{-1/2}$ at cryogenic temperatures.
 The best membrane devices we tested our in a setup without a scanning stage featured $\SF^{1/2}=37\,\mathrm{aN\,Hz^{-1/2}}$~\cite{Catalini_2020}, which is close to the $\SF^{1/2}=19.5\,\mathrm{aN\,Hz^{-1/2}}$ reported at room temperature for trampoline geometries~\cite{Reinhardt_2016}.

Looking forward, we expect to improve the sensitivity by engineering central defects with lower mode masses and better isolation (higher $Q$), as well as by moving to cryogenic operation.
A prospective device will have a mode mass of $77~$pg, a spring constant of $3~$N/m, a frequency of $1\,$MHz, and a room-temperature quality factor of about $1.4\times 10^8$, corresponding to $\SF^{1/2} = 7.6\,\mathrm{aN\,Hz^{-1/2}}$. When cooled to $T=0.2\,$K, we expect $Q = 4.8\times 10^8$ and $\SF^{1/2} = 0.1\,\mathrm{aN\,Hz^{-1/2}}$ (SI), which would be sufficient to detect the magnetic moment of a single proton \cite{Tao_2014}.

An important obstacle for cryogenic detection is the limited displacement sensitivity of our present optical interferometer. The noise in the detected photocurrent amounts to an imprecision in the membrane position measurement of the order $S_{\mathrm{det}}\sim 10^{-28}\,\mathrm{m^2/Hz}$.  When cooled to $T=0.2~$K, the thermal motion of the resonator shown in Fig.~\ref{fig:fig2} is only $x_\mathrm{rms}=5\ee{-14}~$m.  Allowing for some detection bandwidth of the order of tens of Hz, a displacement sensitivity $S_{\mathrm{det}}\lesssim 10^{-30}\,\mathrm{m^2/Hz}$ will be needed (see SI).
A future version of our inverted scanning microscope will therefore incorporate an optical readout cavity in a membrane-in-the-middle geometry~\cite{Jayich2008}.  A comparably low finesse of $\mathcal{F}\approx 500$ will be sufficient to achieve a sensitivity below $10^{-30}~$m$^2$/Hz.

When implementing such optical detection on mechanical sensors with very low thermomechanical force noise, it can become necessary to take the quantum backaction of the optical measurement into account~\cite{Teufel_2009, Anetsberger_2009, mason_continuous_2019}.
This backaction appears as an additional force noise $S_{\rm{qba}}$, produced by the shot noise of the radiation pressure acting on the membrane.
It is bound by a Heisenberg-type relation to $S_{\rm{qba}} {\geq} \hbar^2/S_\mathrm{det}$, where $\hbar$ is the reduced Planck constant.
%
%
For the prospective membrane described above, one reaches a backaction-dominated regime with $S_{\rm{qba}} \geq S_{\rm{th}}$ for $S_\mathrm{det} \approx 10^{-30}~$m$^2$/Hz.
Eventually, one could use advanced quantum measurement protocols to evade quantum backaction \cite{mason_continuous_2019, Brunelli_2020}.


Finally, from our measurements in Fig.~\ref{fig:fig2}(d-f), we conclude that non-contact friction is not a critical obstacle for reaching sub-aN force sensitivity. The non-contact friction observed at $d=1~$nm distance ($\gamma_{\mathrm{nc}} = 5.5\times 10^{-14}~$kg/s) corresponds to a thermal force noise of $S_{\rm{th}}^{1/2} = 0.8\,\mathrm{aN\,Hz^{-1/2}}$ at $T=0.2\,$K.
We expect that the non-contact friction will be further reduced at cryogenic temperatures and larger stand-off distances.  For example, Stipe \textit{et al.}~\cite{Stipe_2001} found a reduction in $\gamma_{\mathrm{nc}}$ by a factor of $6\times$ when lowering the temperature from $295~$K to $77~$K and a reduction $\propto d^{-1.3}$ with distance for a pendulum cantilever.  An even stronger reduction $\propto d^{-3}$ was reported for a parallel cantilever orientation~\cite{Gotsmann2001}.  These results suggest that non-contact friction will be negligible for distances larger than a few nm.

\section{Summary}

In summary, we demonstrate an inverted approach to scanning probe microscopy based on a vibrating silicon-nitride membrane.  Our platform offers excellent sensitivity, approximately $280~\mathrm{aN\,Hz^{-1/2}}$ at room temperature, and a prospect for significant future improvement by optimizing the design of the phononically shielded membranes.  The force sensitivity remains essentially unchanged down to tip-surface separations of $\sim 1~$nm.  We demonstrate topographic imaging of gold nanospheres and tobacco mosaic virus particles dispersed on the membrane surface.

The low force noise and compatibility of membranes with high-finesse optical cavities are expected to permit quantum-limited force detection.  At sub-Kelvin temperatures, prospective devices will have quantum coherence times on the order of $\tau_{coh} = \hbar Q/k_B T \approx 17~$ms, much longer than any other scanning force probe demonstrated to date, as well as force sensitivities around $0.1\,\mathrm{aN\,Hz^{-1/2}}$.
Important applications of ultrasensitive, quantum-limited force microscopy include the investigation of single spins~\cite{Rugar_2004}, atomic-scale nuclear magnetic resonance~\cite{Sidles_1991,Degen_2009}, electrical quantum dots~\cite{Hanson_2007}, or skyrmions~\cite{Meng_2019}, as well as strong spin-mechanics coupling in quantum hybrid devices.

\section{acknowledgement}

We gratefully acknowledge technical support by the mechanical workshop and the engineering office at ETH Zurich.
This work was supported by Swiss National Science Foundation (SNFS) through the National Center of Competence in Research in Quantum Science and Technology (NCCR QSIT) the Sinergia grant (CRSII5\_177198/1) and the project Grant 200020-178863, Danmarks Grundforskningsfond (DNRF) Center of Excellence Hy-Q, the European Research Council through the ERC Starting Grants ``NANOMRI'' (grant agreement 309301), ``Q-CEOM'' (grant agreement 638765), ERC proof-of-concept grant ``ULTRAFORS'' (grant agreement 825797), and a Marie-Curie Fellowship ``Nano-MRI'' (325866), as well as an ETH research grant (ETH-03 16-1).

\providecommand{\noopsort}[1]{}\providecommand{\singleletter}[1]{#1}%

\clearpage

\onecolumngrid
\appendix
\renewcommand{\thepage}{S\arabic{page}}  
\renewcommand{\thesection}{S\arabic{section}}   
\renewcommand{\thetable}{S\arabic{table}}   
\renewcommand{\thefigure}{S\arabic{figure}}
\setcounter{figure}{0}
\setcounter{page}{1}

\large
\begin{center}

\textbf{\textbf{Supporting Information for:} Membrane-based scanning force microscopy}

\normalsize

\vspace{5 mm}

David H\"alg$^{1,*}$, Thomas Gisler$^{1,*}$, Yeghishe Tsaturyan$^{2,3}$, Letizia Catalini$^{2,4}$, Urs Grob$^1$, Marc-Dominik Krass$^1$, Martin H\'eritier$^1$, Hinrich Mattiat$^5$, Ann-Katrin Thamm$^1$, Romana Schirhagl$^6$, Eric C. Langman$^{2,4}$, Albert Schliesser$^{2,4}$, Christian L. Degen$^1$, Alexander Eichler$^{1}$\\

\vspace{5 mm}

\textit{$^1$\affilETH}\\
\textit{$^2$\affilNBI}\\
\textit{$^3$\affilChicago}\\
\textit{$^4$\affilNBIHYQ}\\
\textit{$^5$\affilUnibas}\\
\textit{$^6$\affilGroningen}\\
\textit{$^*$These authors contributed equally to this work.}

\end{center}

\section{Setup}

\begin{figure*} 
    \includegraphics[width=\textwidth]{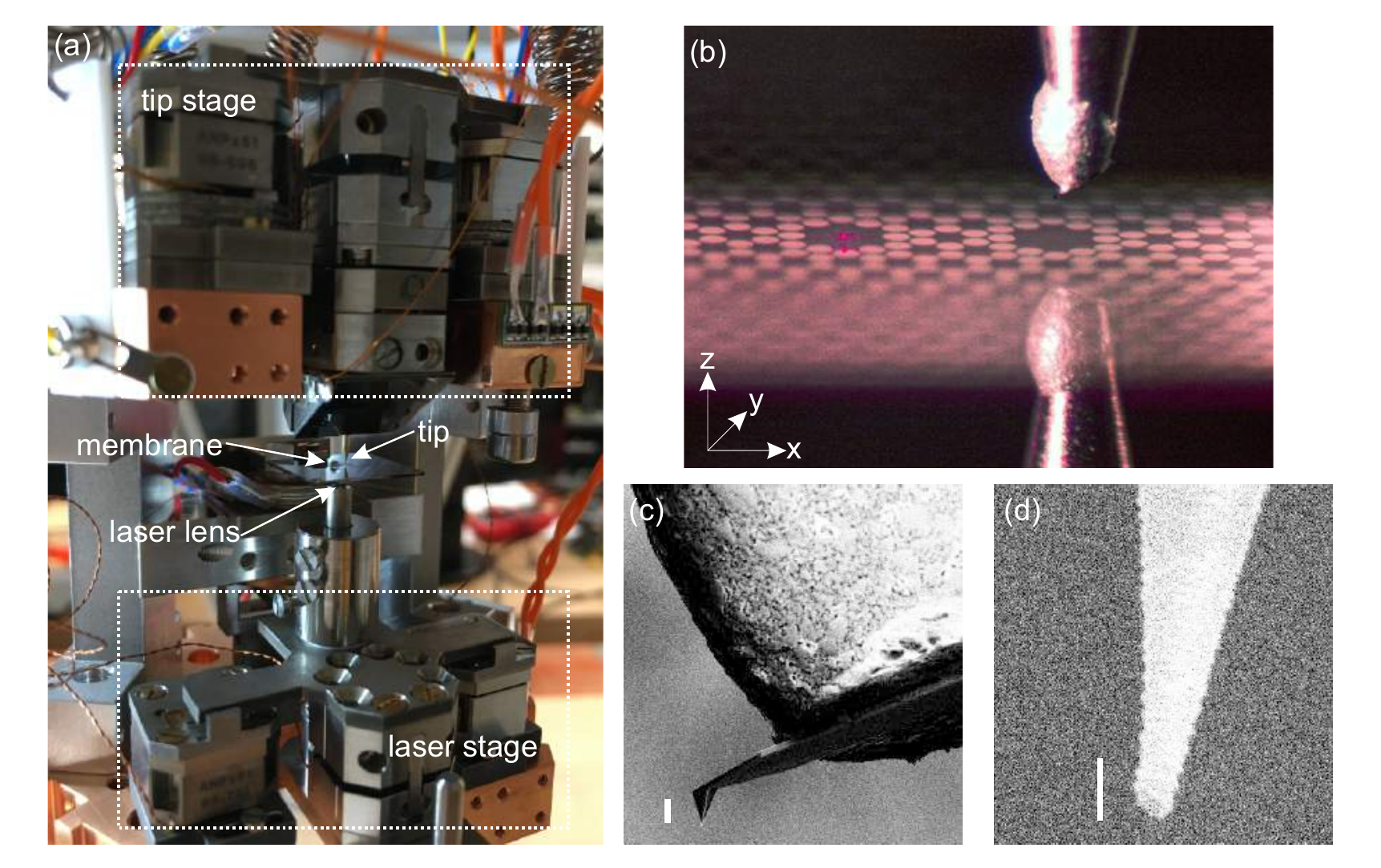} 
    \caption{\textbf{Components of scanning force microscopy stage.} (a) Custom microscope setup with scanning motors to move the tip (top stage) and the laser spot (bottom stage). (b) Optical microscopy image of a membrane mounted in the setup. The left defect mode is used for readout, while the right defect mode serves as a sample plate. (c)-(d) Scanning electron micrographs of the scanning tip used in the main text with scale bars of $10~\mu$m in (c) and $100~$nm in (d).}
    \label{fig:SISetup}
    \end{figure*}

Our instrument comprises a silicon nitride membrane, a commercial atomic force microscope (AFM) tip on a stepping and scanning stage, and an optical interferometer that is focused onto one of the membrane defects (Fig.~\ref{fig:SISetup}).  The laser spot is positioned on the mode with the help of a second motor stage, an optical microscope, and a visible $633~$nm laser (see red spot in Fig.~\ref{fig:SISetup}(b)). We center the interferometer on the defect in x and y before adjusting the lens-membrane distance to match the focal distance of the lens. 

The scanning probe, shown in Fig.~\ref{fig:SISetup}(c) and (d), is a commercial AFM tip (\textsc{OPUS 240AC-PP}) that is glued onto a custom metal needle with a conducting epoxy (\textsc{EPO-TEK H20E}). After the epoxy is cured, we break off the cantilever support chip to avoid unwanted surface interaction.

In addition to the laser and tip driving methods described in the main text, the membrane can be driven by a dither piezo embedded into the sample holder. All data shown in the SI was recorded with this piezo element as a drive.

\section{Sample preparation}

\begin{figure*} 
    \includegraphics[width=\textwidth]{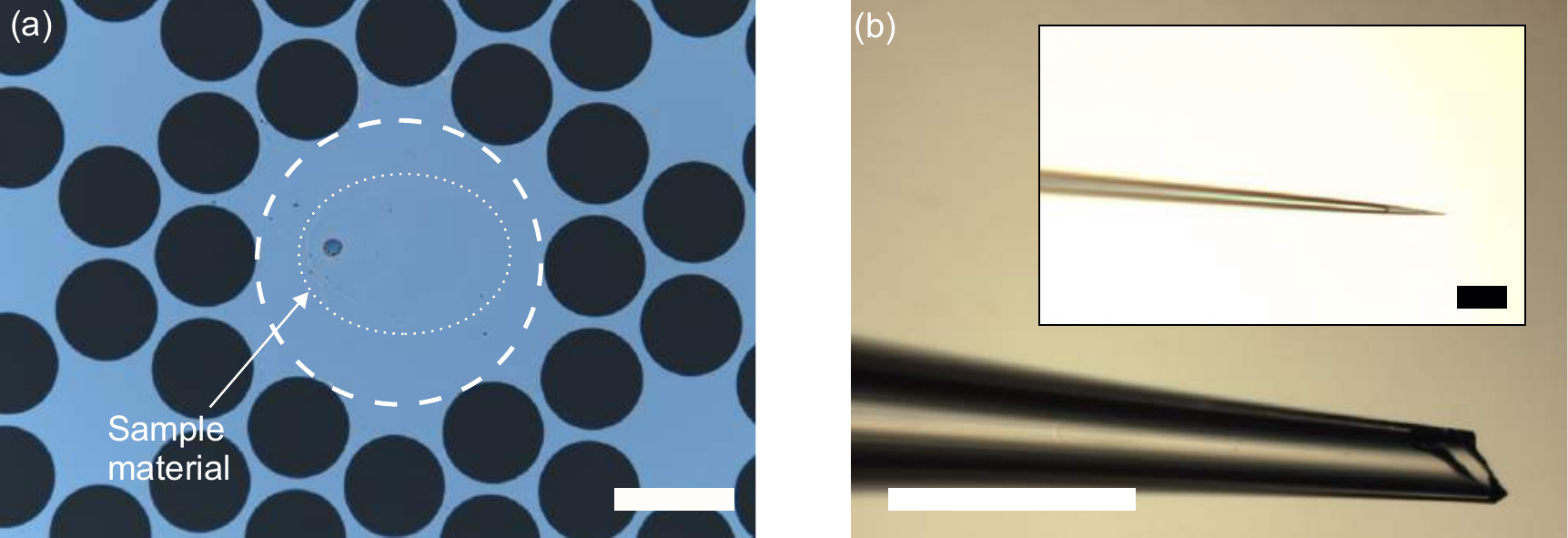} 
    \caption{\textbf{Sample Preparation.} Optical microscopy image (a) showing the membrane area with deposited samples. The image was taken with the double-defect membrane used for this work. The dashed line indicates the membrane defect, while the dotted line emphasises the outline of the deposited sample material. The lateral scale bar is $100~\mu$m. (b) Glass capillary after and before (inset) breaking the tip off. The lateral scale bars are $50~\mu$m.}
    \label{fig:SISempPrep}
    \end{figure*}

The large area of the membrane defect allows to deposit samples from a liquid suspension through a micropipette attached to a micromanipulator. Preceding the sample deposition, the membrane is cured with an UV Ozone Cleaner to eliminate organic contaminants and to increase the hydrophilicity of its surface. The custom micropipette consists of a hollow glass capillary that is sharpened with a \textsc{Narishige} PC-10 puller (see inset Fig.~\ref{fig:SISempPrep}(b)) and connected to a syringe. To achieve a sufficiently high flow, we manually break the capillary tip off.

The syringe allows a controlled deposition of a suspension droplet with samples. Due to residues in the solvents, evaporated droplets leave an unwanted crystalline structures on the membrane. This can be partially avoided by retracting the solvent little by little back into the pipette. Experience has shown that a total impact time of 60 seconds before removing the residual droplet is an appropriate compromise between the deposited amount of sample material and contamination. The membranes are surprisingly robust; however, the risk of structural damage increases with decreasing membrane thickness. The rate of a successful deposition that leaves the membrane intact varies from about 50 percent (for a thickness below $20~$nm) to almost 100 percent ($60~$nm).

We have found that membranes treated in this way preserved quality factors up to $70$ millions, whereas untreated membranes achieved up to $100$ millions in our scanning force setup. The difference could be due to statistical spread, and we have no systematic evidence for a reduction of $Q$ with the sample deposition.

For the work shown in the main text, we used two different types of samples simultaneously: commercially available gold nanoparticles (\textsc{BBI Solutions}) with a nominal mean diameter of 50 nanometers and tobacco mosaic viruses (TMV) cultivated within the group of Prof. C. Degen at ETH Zurich. The two sample types were applied consecutively (first gold, second TMV) with the procedure described above.

In future studies, the liquid dispersion technique may be improved by utilizing femtoliter droplets deposited with a hollow cantilever `nanopipette', which may allow targeted placement of individual macromolecules. Dispersion of particles from gas with a nozzle could preserve even higher quality factors. Finally, shock freezing and drying by sublimation of biological samples is often used to avoid structural damage to the samples incurred through capillary forces during a conventional evaporation process. In our microscope, samples could be freeze-dried on a microscopic sample holder (such as a patterned silicon platelet) and then deposited on the membrane surface with a micromanipulator. We therefore see the potential for studying a wide range of samples with our membrane-based inverted scanning force microscope.

\section{Force sensitivity of an unloaded single-defect membrane}

In the following, we present approach data from a second membrane without samples on the surface. This is a single-defect membrane, such that the scanning tip and laser interferometer spot had to be aligned carefully next to each other. Here we used a scanning tunnelling microscopy (STM) tip made from tungsten by electrochemical wet etching as a scanning probe.

\begin{figure*} 
    \includegraphics[width=0.6\textwidth]{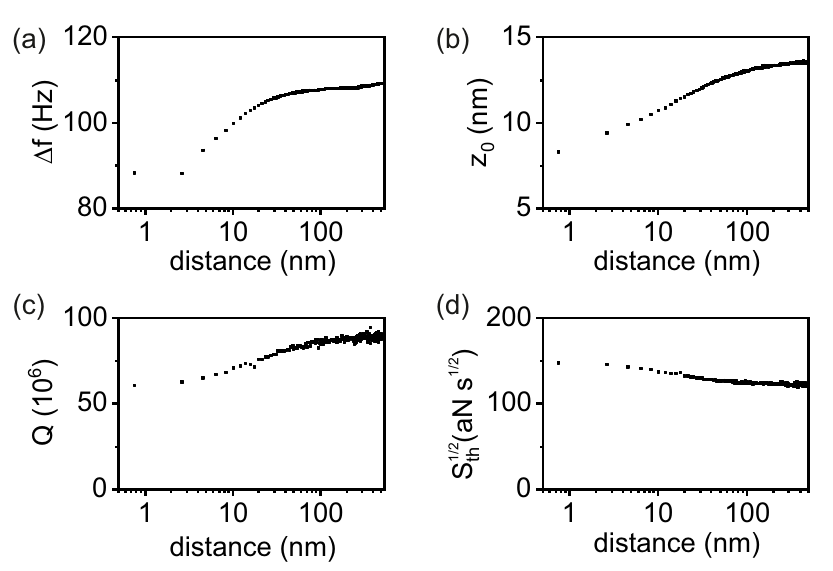} 
    \caption{\textbf{Tip approach with a single defect membrane and a tungsten tip.} (a)-(c) Shift of the measurement frequency, vibration amplitude, and quality factor measured as a function of tip-surface distance, respectively. The membrane is driven by a piezo element and has no sample loaded. (d) Intrinsic thermomechanical force noise calculated from the measured quantities with an effective mass of $m = 16~$ng~\cite{Tsaturyan_2017}.}
    \label{fig:SIApproach}
    \end{figure*}

As we can see in Fig.~\ref{fig:SIApproach}, the membrane retains its excellent force sensitivity in proximity to a scanning tip. An increase of the force noise can only be noticed below $50~$nm distance, and the force noise stays at a very low level even for distance within the nanometer range.

\begin{figure*} 
    \includegraphics[width=0.7\textwidth]{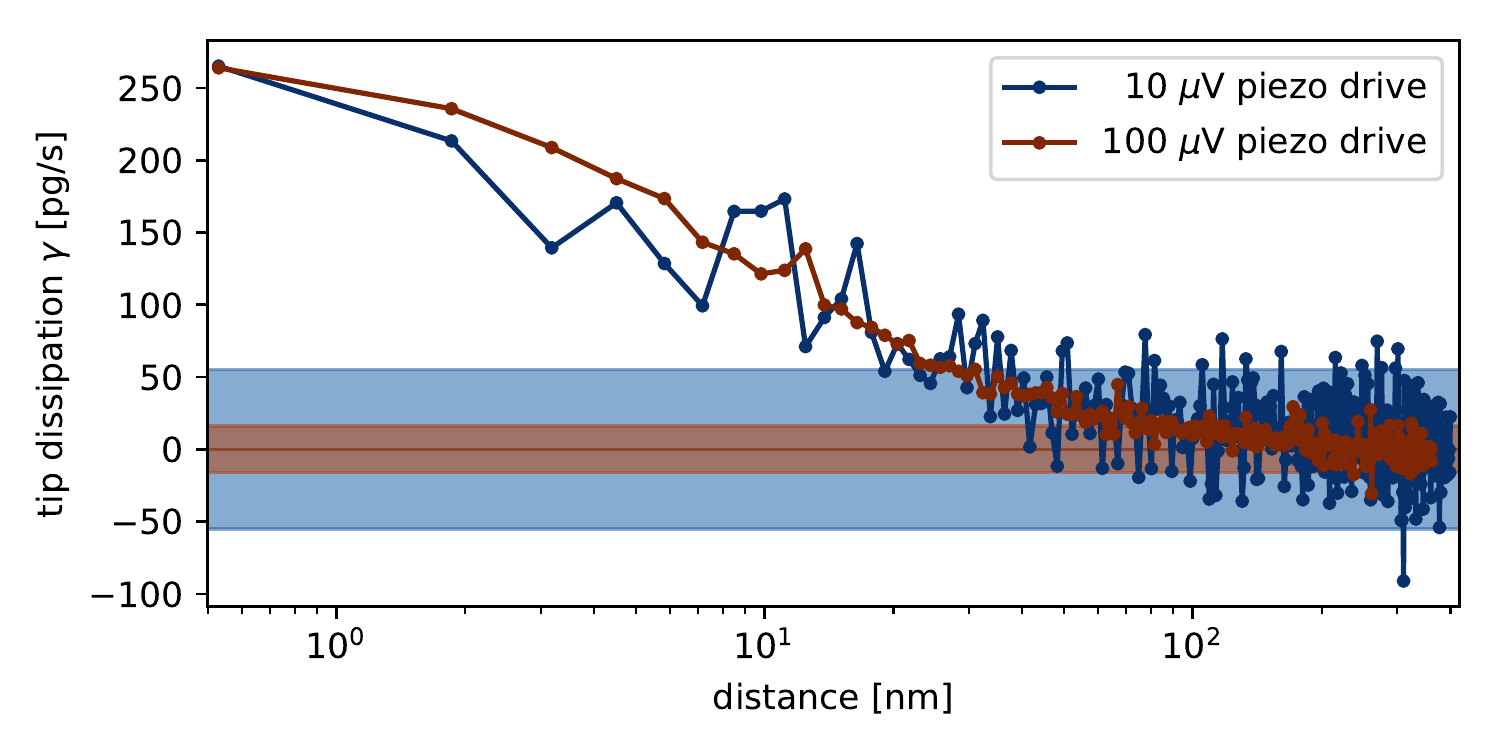} 
    \caption{\textbf{Dissipation induced by the presence of a scanning tip.} Data are calculated from ringdown measurements performed at different distances with two different piezo driving strengths. The average dissipation for large tip-surface distances is used as a baseline, and shaded areas mark the standard deviation of the corresponding uncertainty. These measurements were performed with the same membrane and tip as in Fig.~\ref{fig:SIApproach}.}
    \label{fig:SIDissipation}
    \end{figure*}

In Fig.~\ref{fig:SIDissipation}, we show the tip-sample non-contact friction calculated from the data in Fig.~\ref{fig:SIApproach} (brown) and for a second data set with a drive of $10~\mu$V (blue). Importantly, the data obtained with the two different driving strengths coincide, demonstrating that the dissipation is invariant with the driving force and the vibration amplitude.

    \begin{figure*} 
    \includegraphics[width=\textwidth]{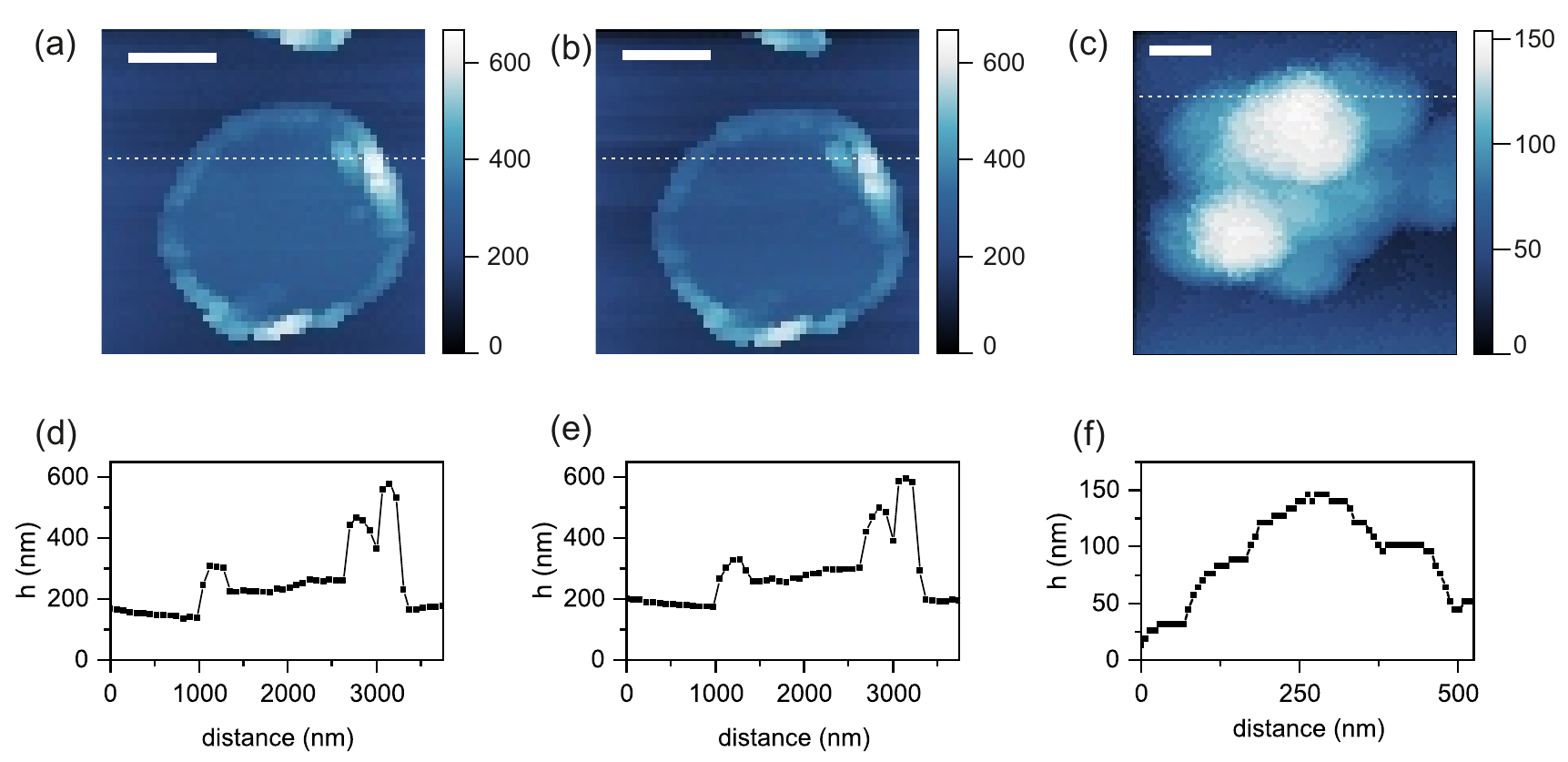} 
    \caption{\textbf{Topography images.} Scans of objects on single-defect membranes. Measurements were performed with piezo driving and with an STM tip made from tungsten. (a) Scan of a gold pad that was evaporated onto the membrane with a nominal height of $100~$nm. Lateral scale bar is $1~\mu$m. (b) Repetition of the scan in (a) to demonstrate the image reproducibility and low drift. No drift compensation was used for the entire measurement time of $26~$hours. (c) Scan of a cluster of gold nanoparticles. Lateral scale bar is $100~$nm. (d)-(f) Linescans extracted from (a)-(c), respectively, at the positions indicated by the dashed lines.}
    \label{fig:SIScan}
    \end{figure*}

\section{Scan data processing and additional scan data}
The scans data in Fig.~3 of the main text and Fig.~\ref{fig:SIScan} were plane-leveled and Fig.~3(a) was additionally line averaged. The scale of the x and y coordinates were calculated with the piezo specifications from the \textsc{attocube} data sheet. The measured TMV sample were used to calibrate the z movement of the \textsc{attocube} scanner.

In Fig.~\ref{fig:SIScan} we show additional topography images obtained with a single-defect membrane and an etched tungsten tip.

\section{Engineering of Low Mass Defects}

\begin{figure}
    \centering
    \includegraphics[width=0.68\linewidth]{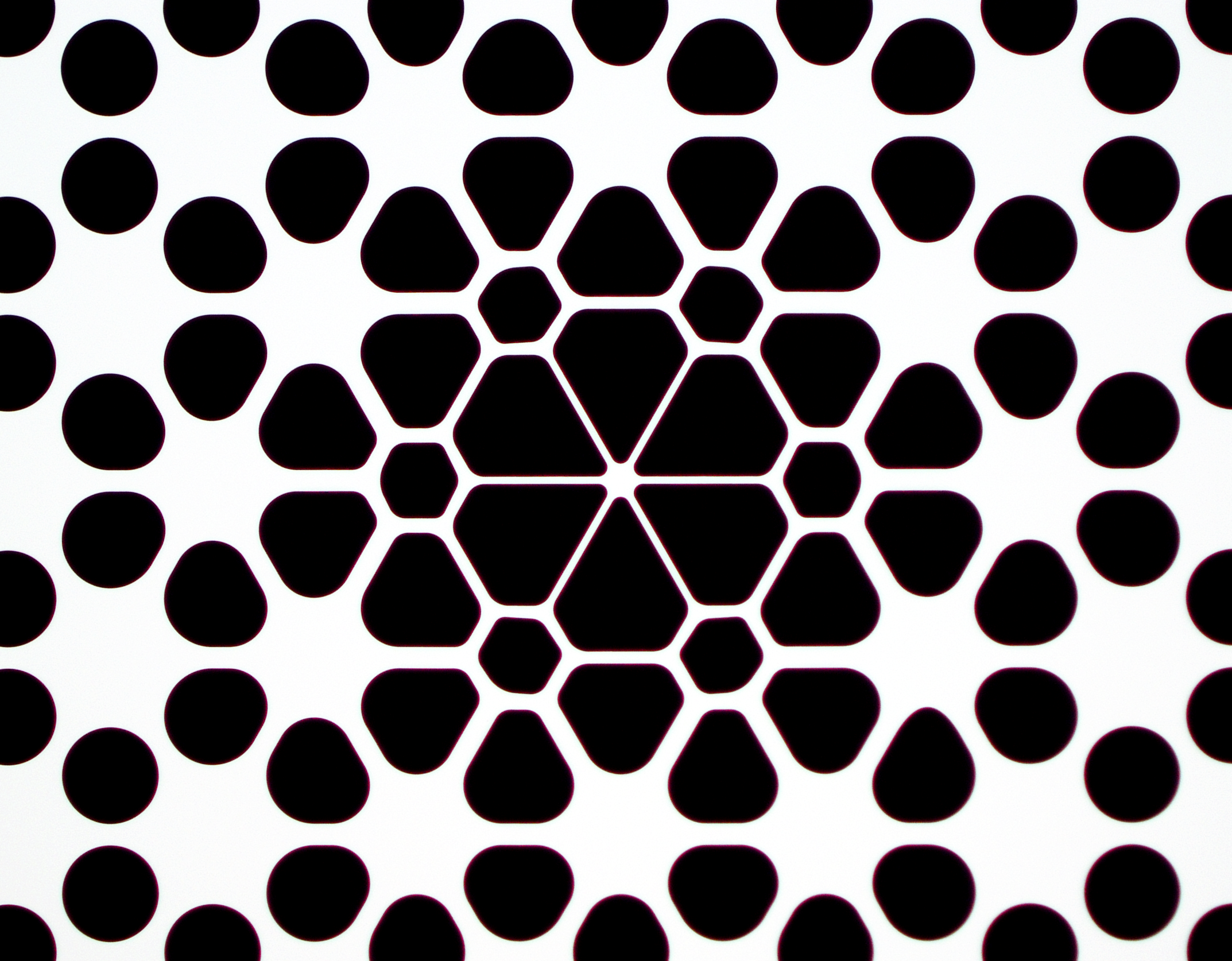}
    \caption{Micrograph image of low-mass defect design being explored for enhanced force sensitiviy. The critical-dimension trampoline tethers are enlarged here, relative to the specific structures analyzed in Table \ref{tab:lowmass}. }
    \label{fig:lowmass}
\end{figure}

\newcommand*{\ra}[1]{\renewcommand*{\arraystretch}{#1}}
\begin{table}[t!]
\centering
\ra{1.3}
\begin{tabularx}{0.95\linewidth}{@{}l|cc|cc|ccc@{}}\toprule

 & \hspace{.2pc} m & k \hspace{.2pc} & \multicolumn{2}{c|}{$Q$ ($\times 10^6$)}  & \multicolumn{2}{c}{ \hspace{.4pc}
 $S^{1/2}_{\mathrm{th}}$  (aN/$\sqrt{Hz}$)}   \\

Design (thickness, frequency) & (pg) & (N/m) &
\hspace{.05pc} 300~K  \hspace{.05pc}  & \hspace{.05pc}   $\leq$ 10~K \hspace{.05pc} &  
\hspace{.5pc} 300~K \hspace{.5pc} & 4~K \hspace{.5pc} & 0.2~K \hspace{.4pc}  

\\ \botrule

Fabricated (50 nm, 1.6 MHz) & & & & & & & \\ 

\hspace{.4pc} Standard  &
8000 & 809  & 58.0   & ---  & 150 & --- & --- \\
\hspace{.4pc} Low Mass: \SI{3}{\micro\meter} &
600 & 61  & 41.5   & ---  & 49 & --- & --- \\
\hspace{.4pc} Low Mass: \SI{1}{\micro\meter}  &
100 & 10 & 15.9   & ---  & 32 & --- & --- \\ \botrule

Rescaled (100 nm, 1.6 MHz) & & & & & & & \\ 

\hspace{.4pc} Standard  &
16000 & 1620 & 29.0   & 120  & 300 & 18.6 & 2.08 \\
\hspace{.4pc} Low Mass: \SI{3}{\micro\meter} &
1200 & 121 & 21.3   & 74  & 97 & 6.0 & 1.35 \\
\hspace{.4pc} Low Mass: \SI{1}{\micro\meter}  &
200 & 20 & 8.0   & 28  & 65 & 4.0 & 0.89 \\ \botrule

Rescaled (15 nm, 1.0 MHz) & & & & & & & \\ 

\hspace{.4pc} Standard  &
6000 & 237 & 470   & 1650  & 36.3 & 2.24 & 0.50 \\
\hspace{.4pc} Low Mass: \SI{3}{\micro\meter} &
460 & 18 & 360   & 1250  & 11.5 & 0.71 & 0.16 \\
\hspace{.4pc} Low Mass: \SI{1}{\micro\meter}  &
77 & 3 & 140 & 480 & 7.54 & 0.47 & 0.10 \\
\botrule
\end{tabularx}

    \caption{Comparing mass $m$, spring constant $k$, quality factor $Q$, and the single-sided force sensitivity of new designs to the standard defect design utilized in previous works~\cite{Tsaturyan_2017,Rossi_2018,mason_continuous_2019, Chen_2020}. For `Low Mass' designs, the critical dimension (\SI{1}{\micro\meter}, \SI{3}{\micro\meter}) corresponds to the width of tethers of the central trampoline defect.}
    \label{tab:lowmass}
    
\end{table}

New phononic crystal/defect combinations are actively being designed and fabricated to increase $Q/m$ and proportionally decrease the force noise. A fabricated single-defect membrane is presented in Fig.~\ref{fig:lowmass}, demonstrating the principle structure behind one such new design. The phononic crystal is tapered down to a six-point trampoline resonator, engineered in such a way as to avoid extremes in the tensile stress in the connecting region. Conveniently, this design also reduces or eliminates non-essential mechanical modes within the bandgap and allows for the primary defect mode to be centered within the bandgap itself. 

In Table~\ref{tab:lowmass}, we compare the $Q$-factor and force sensitivity of the `standard' design to a pair of new `low-mass' designs defined by a critical dimension: the tether width of the trampoline arms. This critical dimension is what predominately determines the mode mass of the resonance, with its lower bound set by the processing steps of the fabrication resolution. The top section of the table combines measured $Q$-factors with simulated mode masses for initial fabrications made with $50~$nm thick membranes. To obtain the rescaled values shown in the two bottom sections of the chart, standard scalings for soft-clamped membranes are used, namely $Q \propto (\mathrm{thickness})^{-1} \times (\mathrm{frequency})^{-2}$. Additionally, it is assumed that the quality factors will increase by a factor $3.5$ when cooling from room temperature to cryogenic temperatures, as observed with previous devices. 

The rescaled values for $1.6~$MHz defect modes fabricated with $100~$nm membranes are directly comparable to the survey results of typical designs conducted by Reetz et al.~\cite{Reetz_2019}.  Our findings suggest a potential for significant improvement in force sensitivity over previous designs, with still a large parameter space to explore for optimization. With operational membranes expected to be $15~$nm thick (or less), the final third of Table~\ref{tab:lowmass} provides optimistic predictions for state-of-the-art membranes utilizing these designs.

\section{Role of displacement imprecision at finite frequency offset}

\begin{figure}
        \includegraphics[width=0.68\linewidth]{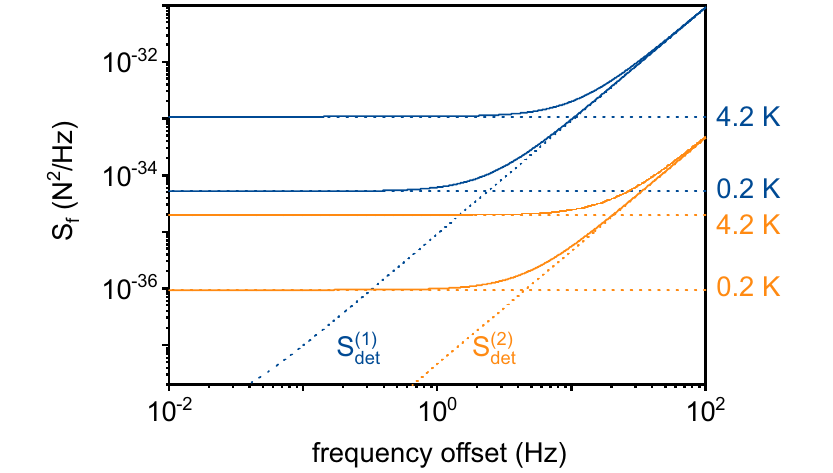} 
        \caption{\textbf{Effective force noise PSD.} Calculated apparent force noise $S_f = S_{\rm{th}} + S_{\rm{det}}/|\chi_\mathrm{m}(f)|^2$ consisting of thermomechanical force noise $S_{\rm{th}}$ (horizontal dotted lines) and displacement imprecision $S_{\rm{det}}$ (diagonal dotted lines), where $\chi_\mathrm{m}(f)$ is the mechanical susceptibilty. Blue solid lines correspond to the predicted performance of the device used in this work at $4.2~$K and $0.2~$K with $S_{\rm{det}}^{(1)} = 2\times 10^{-28}~$m$^2/$Hz (single-sided convention). Orange solid lines are calculated for a device tested without force scanning, with $m = 2~$ng, $f_0 = 1.35~$MHz, and $Q = 2\times 10^8$~\cite{Catalini_2020}. For this device, we project the numbers obtained when using an optical cavity with $S_{\rm{det}}^{(2)} = 10^{-30}~$m$^2/$Hz for readout.
        }
        \label{fig:readout_noise}
        \end{figure}

In Fig.~\ref{fig:readout_noise}, we show the impact of the displacement imprecision on the effective force noise of the membrane sensor s used in this work as a function of frequency offset (from $f_0$).
This becomes important, for instance, for the detection of nuclear spin signals with lifetime-limited spectral widths (e.g. $50~$Hz for hydrogen signals in a typical NanoMRI measurement).
For the parameters of this sensor, quantum backaction effects are not yet relevant.

\section{Growth of tobacco mosaic viruses}

Tobacco mosaic viruses (TMV) were produced by infected tobacco plants~\cite{Birnbaumer_2009}. Young plants are grown from the seeds and can be easily infected when they have approximately 4 leaves. At this stage, they are strong enough to survive the treatment but not to withstand the virus infection. Infecting is done by rubbing a small area on a leaf with fine sand and dropping the virus solution on the injured area. The success of infection can be confirmed by optical examination of the leaves after a few weeks. Leaves on plants suffering from the mosaic sickness tend to be smaller and have a mosaic structure consisting of light and dark green patches. The infected plants are then grown for several months until they are ready to be harvested. In order to extract the viruses, the vein-free parts of infected tobacco leaves are crushed in liquid nitrogen and stored for further usage at $-80^{\circ}~$C. The dried powder is re-suspended in buffer containing 2-mercaptoethanol. The mixture is filtered through a large-pored cheese cloth to remove large parts. Then, butanol is added to the filtrate and stirred, followed by centrifugation. The supernatant is removed and NaCl and poly(ethylene glycol) are added to the suspension to precipitate the proteins. The suspension is centrifuged again and the supernatant is discarded. The pellet is re-suspended in KH2PO4/K2HPO4 buffer containing Triton X-100. Finally, a density gradient centrifugation is performed using a sucrose pad (buffered $25~\%$ percent sucrose). The pellet is washed and centrifuged until the required purity is obtained. The protein concentration can be determined by UV-visible spectroscopy. The success of the purification process was confirmed by observing the characteristic rod-shaped viruses under a scanning electron microscope and a conventional atomic force microscope.

\providecommand{\noopsort}[1]{}\providecommand{\singleletter}[1]{#1}%

\end{document}